\documentclass[fleqn,usenatbib]{mnras}

\usepackage{newtxtext,newtxmath}

\usepackage[T1]{fontenc}
\usepackage{ae,aecompl}

\usepackage{graphicx}   
\usepackage{amsmath}   
\usepackage{amssymb}   

\newcommand{\fagn}{f_{\rm AGN}}
\newcommand{\fSFG}{f_{\rm SFG}}

\newcommand{\ana}{A\&A}

\newcommand{\kms}{{\rm~km~s^{-1}}}

\newcommand{\SFRf}{\rm SFR_{fib}}

\newcommand{\SFRfib}{\rm SFR_{fib}}

\newcommand{\fbar}{\mathinner{f_\mathrm{bar}}}

\title[Bars and Galaxy Secular Evolution]
{Effect of Bars on Evolution of SDSS Spiral Galaxies}

\author[M. Kim et al.]{
Minbae Kim,$^{1}$
Yun-Young Choi,$^{1,2}$
and Sungsoo S. Kim$^{1,2}$
\\
$^{1}$School of Space Research, Kyung Hee University, Yongin-shi, Kyungki-do 446-701, Republic of Korea \\
$^{2}$Department of Astronomy and Space Science, Kyung Hee University, Yongin-shi, Kyungki-do 446-701, Republic of Korea
}

\date{Accepted XXX. Received YYY; in original form ZZZ}

\pubyear{2019}

\begin{document}
\label{firstpage}
\pagerange{\pageref{firstpage}--\pageref{lastpage}}
\maketitle

\begin{abstract}
We explore the significance of bars in triggering central star formation (SF) and AGN activity
for spiral galaxy evolution using a volume-limited sample with 
$0.020<z<0.055$, $M_{\rm r}<-19.5$, and $\sigma>70\kms$ selected from SDSS DR7. 
On a central SF rate-$\sigma$ plane, we measure the fraction of galaxies with strong bars
in our sample and also the AGN fractions for barred and non-barred galaxies, respectively.
The comparison between the bar and AGN fractions
reveals a causal connection between the two phenomena of SF quenching and AGN activity. 
A massive BH and abundant gas fuels are sufficient conditions to trigger AGNs. 
We infer that the AGNs triggered by satisfying the two conditions 
drive the strong AGN feedback, suddenly suppressing the central SF and leaving 
the SF sequence.
We find that in galaxies where either of the two conditions is not sufficient, 
bars are a great help for the AGN triggering, accelerating 
the entire process of evolution, which is particularly evident in pseudo-bulge galaxies.
All of our findings are obtained only when  
plotted in terms of their central velocity dispersion and central SFR (not galactic scale SFR),
indicating that the AGN-driven SF quenching is confined in the central kpc region.
\end{abstract}




\section{Introduction}
\label{sec:sec1}

Gas-inflows into the galactic central region play an important role in galaxy formation and evolution.
Understanding how to drive gas into the small scales of galactic nuclei has been an important issue 
over the past two decades, and
various gas-inflow mechanisms have been proposed; for example,
on galactic scales, galaxy-to-galaxy interaction, major mergers 
\citep{Sanders1988, Springel2005, Dimatteo2005, Hopkins2006, Alonso2007, Kim2020}, 
minor mergers \citep{Roos1981,Hernquist1995}, 
and bar-driven gas inflow \citep{Combes2003, Kormendy2004} 
and on smaller scales, turbulence of interstellar medium (ISM) in galactic discs 
\citep{Elmegreen1998, Wada2004, Wada2009, Kawakatu2008}, 
stellar wind \citep{Ciotti2007, Davies2012}, and so on.

In this study, we focus on the role of bar-driven gas inflow for triggering nuclear activity of a galaxy.
Gravitational interactions between stars and gases in the galactic bars reduce the angular momentum of gases, 
driving them into a few hundred parsec central region 
 \citep{Lynden1979, Shlosman1990, Sellwood1981, Heller1994, Combes2014, Carles2016}. 
Based on this idea, many simulations have shown that 
the elongated bar structure can drive gas into central regions of the galaxy 
\citep{Friedli1993, Debattista1998, Athanassoula2003}.
The large-scale bar potential forms another non-axisymmetric substructure inside the bar 
that helps to cause the gas inflow further closer to the central BH: 
nuclear spirals \citep{Ann2005, Thakur2009, Kim2017b} 
or nuclear rings \citep{Kim2011, Shin2017} at several hundred parsecs, 
and nuclear bars or secondary bars \citep{Namekata2009} at a few ten parsecs. 

The presence of a large-scale bar often accompanies star formation (SF) and/or AGN. 
Indeed, several observational studies found that, compared to non-barred galaxies,
barred galaxies show a higher central SF activity 
\citep{Heckman1980, Knapen2002, Jogee2005, Hunt2008, Bang2009, Hao2009, Ellison2011, Oh2012, Wang2012, Consolandi2017} 
or a higher AGN fraction 
\citep{Hao2009, Oh2012, Alonso2013, Galloway2015}. 
However, some studies found no compelling evidence for AGN-bar
\citep{Lee2012b, Cheung2015a, Cisternas2015, Goulding2017} 
or SF-bar connection \citep{Martinet1997, Chapelon1999, Willett2015}. 
\citet{Cheung2015b} found in quiescent SDSS spiral galaxies
no difference in the stellar populations and chemical evolution of barred and non-barred galaxies. 
Even the anti-correlations between bar presence and SF activity 
\citep{Cheung2013, Gavazzi2015, Consolandi2017} and between bar presence and HI gas 
\citep{Masters2012, Cervantes2017} were found. 
\citet{Consolandi2017} demonstrated the spatial correlation between 
the presence of a bar and the central cold dust content from the FIR emission.
has also been demonstrated by \citet{Consolandi2017}. 

These confusing and controversial observational results suggest
that SF or AGN is not observed in all barred-galaxies.
That is, nuclear activities are likely to be closely related to bar instability 
rather than the bar itself \citep[e.g.,][]{Fanali2015}.
Bar instability has to do with galaxy properties such as star formation rate (SFR) and
gas availability. Several studies using SDSS spiral galaxies support the point.
\citet{Lee2012b} found that, when colour and mass are fixed, a slight excess of pure AGN 
\citep{Kewley2006} fraction is seen in barred galaxies (but with large uncertainty).
Bars in red spiral galaxies are more frequently seen and stronger than those in blue spiral galaxies 
\citep{Masters2010, Masters2011, Lee2012a, Oh2012, Alonso2013, Cheung2013}, 
which is consistent with a result of \citet{Kim2017a} 
showing that SF activity and gas amount in galaxies with strong bar are lower than those of non-barred ones. 

Positive bar-SF and bar-AGN connections are more common in blue spiral galaxies. 
Therefore,
when investigating the role of bars in the galaxy secular evolution accompanying SF and AGN activities,
the galaxy properties closely related to growth of bar instability should be considered.

In this work, we investigate the relationship between strong-barred spiral galaxies and AGN activities 
using the same morphological classification as \citet{Lee2012a, Lee2012b} 
and a more conservative AGN classification with higher signal-to-noise criteria.
\citet{Lee2012b} found no evidence for an AGN-bar connection even at fixed colour and $\sigma$. 
The biggest difference between our study and that of \citet{Lee2012b} is that we 
pay more attention to the various properties at the galactic centre
that are expected to be more closely related to nuclear activities 
such as bulge prominence, central velocity dispersion, central mass concentration, and central SF 
rather than the measurements from the entire galaxy
such as a colour, total SFR, and stellar mass.

The rest of this paper is organized as follows. 
We describe our sample and AGN selection in Section~\ref{sec:sample} and
show how the bar fraction depends on the star formation rate and velocity dispersion 
at the central region in Section~\ref{sec:fbar}. 
At given galaxy properties, the bar effects on AGN triggering 
and AGN activity level are examined in Section~\ref{sec:AGN}.
Because bar formation and evolution are correlated with central mass concentration of the galaxy, 
we also explore how the bar effect and the scaled bar length vary when the central
properties are determined 
after dividing the sample into two bulge-type samples, presented in Sections~\ref{sec:CMC}
and \ref{sec:length}.
Finally, the discussion is presented in Section~\ref{sec:discuss}.

\section{Data and Sample Selection}
\label{sec:sample}

We use the volume-limited face-on spiral galaxy sample of \citet{Lee2012a,Lee2012b} with
the $r$-band absolute magnitude $\mathrm{M_{r}<-19.5}+5{\rm log}h$ 
(hereafter, we excluded the $+5{\rm log} h$ term in absolute magnitude calculation) 
and redshifts $0.020<z<0.055$, selected from Sloan Digital Sky Survey Data Release 7 
\citep[SDSS DR7;][]{Abazajian2009}. 
Throughout this paper, the cosmological parameters are assumed from the $\Lambda$CDM cosmological model 
using density parameters $\Omega_m=0.27$ and $\Omega_{\Lambda}=0.73$.

\subsection{Morphology Classification}

Morphology classification is adopted from 
the Korea Institute for Advanced Study DR7 Value-Added Galaxy Catalogue \citep[KIAS DR7-VAGC;][]{Choi2010}, 
which is complementary to 
the New York University Value-Added Galaxy Catalogue \citep[NYU VAGC;][]{Blanton2005}. 
Galaxies are well divided into 
early-type (ellipticals and lenticulars) and late-type (spirals and irregulars) 
based on their locations in $u - r$ colour versus $g - i$ colour gradient space 
and $u - r$ colour versus $i$-band concentration index space \citep[see,][]{Park2005}. 
With additional visual inspection by \citet{Lee2012a}, 
some galaxy morphologies are changed. 
In this study, we use only late-type galaxies. 

In \citet{Lee2012a}, 
galaxies are classified into four groups based on visual inspection of SDSS colour images: 
strong-barred, weak-barred, ambiguous-barred,and  non-barred galaxies. 
Classified barred galaxies are in good agreement with classification of \citet{Nair2010}. 
In this paper, we define only strong-barred galaxies as barred galaxies, 
which have a bar size larger than $25\%$ of galaxy size. 
To ensure selection of barred galaxies by visual inspection, 
only galaxies with a ratio between $i$-band isophotal minor and major axis, ${\rm b/a}$,
higher than 0.6 are included.
Non-barred galaxies are defined as galaxies not strong or weak or ambiguous-barred. 

Fibre star formation rate, $\SFRfib$, is obtained from MPA-JHU DR8 catalogue. 
$\SFRfib$ of the star-forming galaxies (SFGs) is estimated from emission lines 
\citep{Brinchmann2004},  
while that of others types is estimated from fibre photometry \citep{Salim2007}. 
 
Fundamental photometry parameters of ${M_{r}}$ and $u - r$ colour 
are adopted from the KIAS DR7-VAGC. 
The $u-r$ colour and  $\mathrm{M_{r}}$ have been de-reddened for Galactic extinction 
\citep{Schlegel1998} and k-corrected to redshift $z = 0.1$ \citep{Blanton2003}. 
For stellar velocity dispersion, $\sigma$ adopted from NYU-VAGC, only spectra with    
mean {\rm S/N} per spectral pixel greater than $10$ are used.
A simple aperture correction of $\sigma$ is made using the formula of \citet{Bernardi2003} 
to correct errors due to the finite size of the optical fibre. 
In this study, we use only galaxies with $\sigma>70 \kms$
to reduce systematic errors induced by the instrumental resolution of SDSS spectroscopy
and to avoid a selection effect due to the [O III] flux limit in detecting AGN.

This $\sigma$ cut excludes many disk-dominated and irregular late-type galaxies 
in our late-type galaxy sample, resulting in a final sample of 6,195 spiral galaxies, 
of which 1,893 (31\%) are barred and 3,754 (61\%) are non-barred. The rest are
weak or ambiguous barred galaxies.

\subsection{AGN Selection}

Type ${\rm II}$ AGNs are separated from SFGs based on the flux ratios of Balmer and ionization lines 
\citep[BPT diagram;][]{Baldwin1981, Veilleux1987}. 
The activity types are classified based on the ratios of emission lines
($\rm H\alpha$, $\rm H\beta$, $[\rm OIII]\lambda5007$, and $[\rm NII]\lambda6584$)
detected with  ${\rm S/N}\geq6$. 
We classify the activity types of galaxies using a conservative AGN definition from \citet{Kewley2006}. 
The pure AGNs located above the maximum starburst line of \citet{Kewley2001} contain
Seyfert and low-ionization nuclear emission-line regions (LINERs).
The starburst-AGN composite galaxies located between the maximum starburst line of \citet{Kewley2001}  
and the pure star-forming line of \citet{Kauffmann2003} contain 
both AGN and extended H{\rm II} regions.
Since optical continuum dominated by non-thermal emissions from Type {\rm I} AGN 
makes it difficult to study their host galaxies,  
we excluded potential Type {\rm I} AGNs
that have a $\rm H\alpha$ emission line width larger than $\sim 500\kms$ (FWHM).  

In this study, we define an AGN host  
by combining composite galaxies and pure AGNs. 
All the LINERs identified by the BPT diagram are not bona-fide AGN. 
Some of the weak LINERs are retired galaxies powered by hot low-mass evolved stars 
rather than low luminosity AGNs 
\citep{Stasinska2008, Cidfernandes2010, Cidfernandes2011, Melnick2013}. 
By adopting a criterion of \citet{Cidfernandes2011},  
we excluded the ambiguous objects with $W_\mathrm{H\alpha}<3\AA$ from pure AGN.
Out of our final sample, 58\% are classified as conservatively-defined emission galaxies
with ${\rm S/N}\geq6$.
The details are listed in Table~\ref{tab:tab1}. 
Table~\ref{tab:tab1} also shows how the activity types of the sample galaxies vary 
depending on a the ${\rm S/N}$ threshold adopted for emission line detection.
As the threshold of ${\rm S/N}$ decreases to 3, commonly used in the BPT-AGN classification, 
both the number of SFGs and composite objects increase by approximately 30\%. 
On the other hand, the number of pure AGNs increases by 12\%, 
and that of retired galaxies that were excluded from pure AGNs increases by 165\%. 
Therefore, the contribution of the retired galaxies to the AGN fraction should be noted
when using the low ${\rm S/N}$ threshold.

\section{Results}

\subsection{Bar Effect on Central Star Formation} 
\label{sec:fbar}

\begin{figure*}
    \centering   
   \includegraphics[scale=0.9]{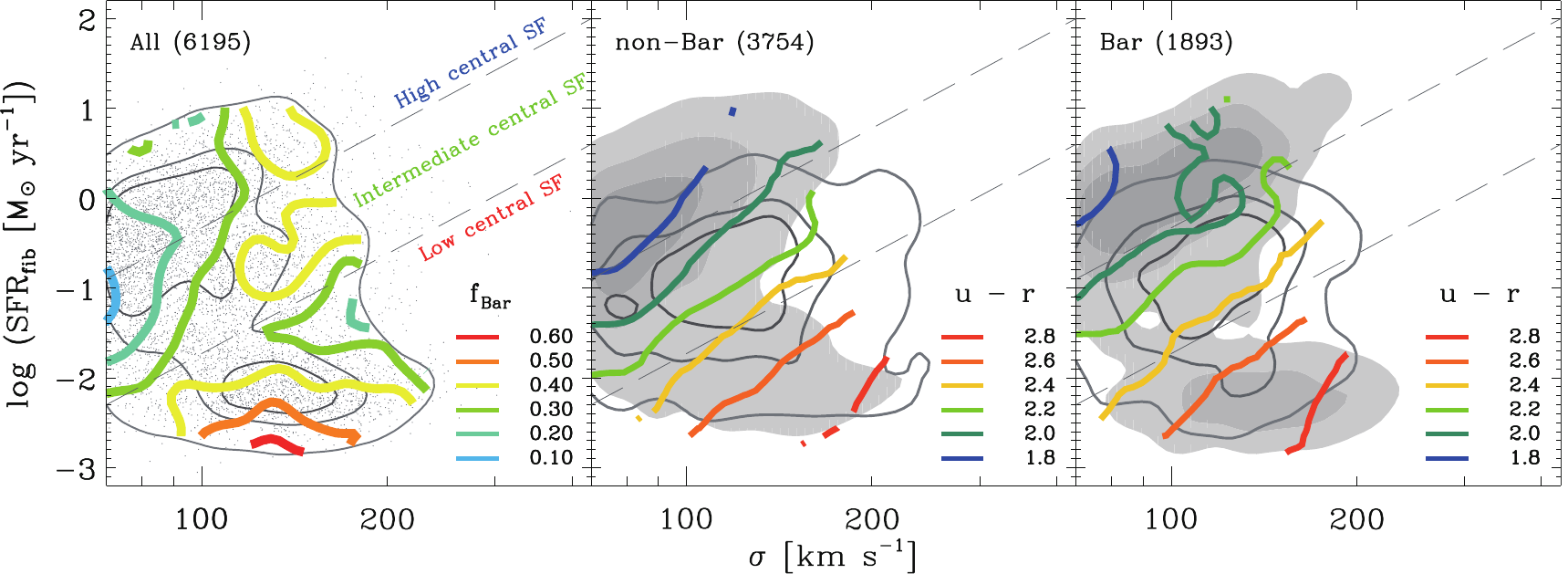}
    \caption{Distributions of bar fraction (left) and median $u-r$ colour of barred (middle)
and non-barred galaxies (right) in $\SFRfib$-$\sigma$ space. 
Each distribution is plotted with coloured thick solid line contours.
The grey thin line contours in the left panel denote constant galaxy number densities.
The grey filled and dark grey line contours in the middle and right panels denote constant number densities 
of non-AGN and AGN host galaxies in each sample, respectively.
Non-AGN galaxies include all those excluding AGNs
defined by emission line ratios without ${\rm S/N}$ cut. 
The bin sizes are $\Delta{\mathrm{log}}\sigma=0.013$ and $\Delta{\mathrm{log}}\SFRfib=0.090$. 
The galaxy density contours enclose 0.5$\sigma$, 1$\sigma$, and 2$\sigma$ of each sample. 
}
    \label{fig:fig1}
\end{figure*}

First, we investigate how physical properties at the galactic centre
are related to the presence of a strong bar.  
This is because bars rearrange the disc gas 
content and drive it to the central region, changing the central properties,
such as central mass concentration (CMC) and central SFR.

We measure a bar fraction, $\fbar$, in two-dimensional space in terms of $\sigma$ and $\SFRfib$.
The $\fbar$ is defined as 
ratio of the number of barred galaxies to the number of total galaxies. 

As a proxy of central SFR within the inner few kpc of a galaxy,
we use a fibre SFR measured within a SDSS fibre aperture, $\SFRfib$. 
At the median redshift of our sample, $z = 0.043$, 
the physical diameter of the fibre aperture corresponds to $\sim 2.7$ kpc, 
which is similar to the average size of bulges in the late-type galaxies we selected 
\citep[e.g.,][]{Fisher2010, Cheung2015b}. 
One can estimate fibre size from the SDSS images as in Figure~\ref{fig:fig9} below.
We also fix central velocity dispersion, $\sigma$, 
because CMC influences $\sigma$ \citep{Athanassoula2005}.

In the left panel of Figure~\ref{fig:fig1},  
coloured thick solid line contours show the distribution of $\fbar$ 
in two-dimensional space in terms of $\SFRfib$ and $\sigma$.
Weak barred and ambiguous galaxies are excluded from our bar selection.
For comparison, the number density distribution of the whole sample is plotted 
with grey thin line contours.
For each of non-barred and barred galaxies, the number density distributions of non-AGN 
and AGN host galaxies are also plotted with grey filled and grey line contours
in the middle and right panels, respectively.

Hereafter, unless otherwise noted, 
all the smoothed distributions in the figures of this study are obtained using the fixed-size 
spline kernel for each bin (60 by 60) in the parameter space explored. 
The significance level for the smoothed galaxy distributions is calculated
using the bootstrap method with 1000 runs. 
Each contour encloses $0.5\sigma$, $1\sigma$, and $2\sigma$ in order.
The uncertainties for the $\fbar$ (or AGN fraction below) measurements 
were also calculated using the bootstrap method. 
For bins with the lowest $\fbar$ of 0.1, the relative uncertainty is less than 15\%.

The notable feature is found in the galaxy number density distributions
in $\SFRfib$ and $\sigma$ space, which exhibits three distinct peaks.
Comparing the middle and right panels, 
we infer that the barred galaxy sample contributes greatly to the distinct trimodal 
distribution feature that appears to result from two relatively rapid SF quenchings.
We also note that the dependency of $\fbar$ on the $\SFRfib$ and $\sigma$ 
changes before and after the rapid quenching processes.

Considering these features,
we divide the parameter space into three locations with two long-dashed diagonal lines,
for convenience. Galaxies in different locations are at different evolutionary stages. 
In the middle and right panels, the distributions of median $u-r$ colour for non-barred
and barred galaxies are overlaid on the same space 
with coloured thick solid line contours, respectively.
The $u-r$ colour is used as an indicator of SF history of a galaxy.
As a galaxy ages, its $u-r$ colour becomes redder.
In Table~\ref{tab:tab2}, 
the demographics for galaxies in each location are shown in detail.


Above the upper diagonal line, there is a sequence of galaxies
having a tight correlation of $\SFRfib$ and $\sigma$.
This includes 32\% of our sample, most (83\%) of which are classified as SFGs. 
Their AGN fraction is only 6\%. 
These galaxies have blue colour (mostly $u-r < 2.0$).
The $\fbar$ in this region strongly depends on $\sigma$;
bars are most frequently found around $\sigma \sim 130\kms$, 
but the number is small.
It is noteworthy that the difference in $u-r$ colour between barred and non-barred galaxies 
is larger in the SF main sequence than in any other region;
barred galaxies are significantly redder and their $u-r$ also strongly depends on $\sigma$.
In other words, at given colour and $\sigma$, barred galaxies exhibit 
significantly higher $\SFRfib$ than non-barred galaxies, suggestive of gas-inflow exerted by a strong torque from bars
\citep{Hunt1999, Jogee2005, Spinoso2017}.

Between the two diagonal lines, there are
galaxies having intermediate colour of about $2.0<u-r<2.4$, and 50\% of them host an AGN. 
When using the low ${\rm S/N}$ threshold, 60\% host an AGN
and 30\% are SFGs.
This is consistent with the previous result that most  
AGNs reside in green valley galaxies \citep{Goulding2010, Schawinski2009}. 
At a given $\sigma$,
AGN hosts and SFGs show very distinct differences in the central SFR. 
The difference in total SFR of the two populations is rather smaller.
We infer from the result that a possible process causing SF quenching seems to be confined 
in the galactic central region.

This is more pronounced in the case of barred galaxies.
For barred galaxies, about 72\% host an AGN, 
almost twice that of non-barred counterparts.
That is, bars promotes the SF quenching process.
The barred galaxy exhibits higher central SFR at a given colour
and is more likely found at lower $\SFRfib$ at a given $\sigma$;
the features are evident in galaxies with $\sigma<130\kms$, 
suggesting that bars effectively accelerate central gas consumption 
as long as there is no further inflow of gas from outside, particularly in small-$\sigma$ galaxies.
It also means that bars can contribute to rapid SF quenching in the 
galactic central region \citep[see,][]{Gavazzi2015, Consolandi2017, Rosas-Guevara2020}. 

Finally, let us focus on low-$\SFRf$ galaxies below the second diagonal line 
having a red colour of $u-r>2.4$. 
Out of the barred galaxies, about 40\% reside in this location, 
while this percentage for non-barred galaxies is 27\%. 
The $\fbar$ distribution shows that
barred galaxies with $\sigma>130\kms$ are suddenly quenched
compared to their non-barred counterparts. 

\subsection{Bar Effect on AGN}\label{sec:AGN}
The previous section indicated that 
barred galaxies evolve more rapidly than non-barred galaxies 
through accelerated gas consumption by bars, 
implying that bars accelerate the SF quenching process in the central region. 
Then, how efficiently do bars lead to inflows to the nuclear region 
while they accelerate starbursts?

\subsubsection{Bar Effect on AGN Fraction}\label{sec:AGN}
To estimate how effectively bars affect AGN, 
we measure a bar effect as 
the ratio between the AGN fractions in the barred and the non-barred galaxies 
on the $\SFRfib$-$\sigma$. 
A ratio greater than 1.0 indicates a positive bar effect on AGN triggering. 
The AGN fraction $\fagn$ is calculated 
as the ratio between AGN hosts and all galaxies. 

\begin{figure*}
    \centering   
   \includegraphics[scale=0.9]{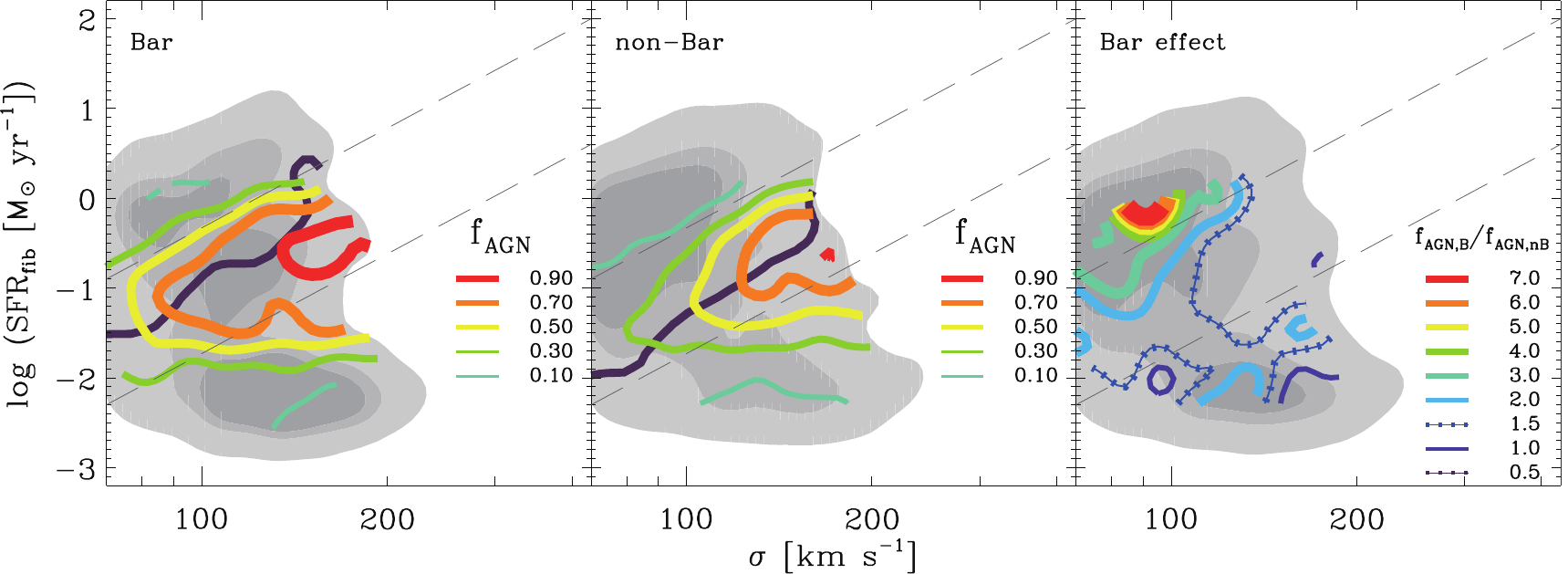}
    \caption{Distributions of AGN fractions of barred (left) and 
non-barred (middle) galaxies in $\SFRfib$-$\sigma$ space. 
Bar effect on AGN triggering (the ratio between the $\fagn$s
 in barred and non-barred galaxies) is plotted with coloured solid lines at the right panel 
and is only calculated when the AGN fractions of barred and non-barred galaxies are both valid.  
A ratio greater than 1.0 indicates a positive bar effect on AGN activity. 
Grey filled contours represent 
the number density distribution of each sample.
Each contour encloses 0.5$\sigma$, 1$\sigma$, and 2$\sigma$ of each sample. 
The violet thick solid line denotes a constant median $u-r$ colour of 2.2 of each sample. 
}
    \label{fig:fig2}
\end{figure*}

The result is plotted with coloured thick solid lines
in the right panel of Figure~\ref{fig:fig2}.
The violet thick solid line denotes a constant median $u-r$ colour of 2.2 of each sample.
Grey filled contours in the left and middle panels show that 
the trimodal number distribution over $\SFRfib$-$\sigma$ space
is more pronounced in barred galaxies. 

The $\fagn$ increases as galaxies leave the SF sequence. 
These results remind us that starbursts and AGNs co-evolve 
\citep[e.g.,][]{Dimatteo2005, Hopkins2005}. 
In barred galaxies, evolution from SFGs to AGNs occurs more abruptly,
leading to more significant increase in $\fagn$ in the intermediate $\SFRf$ location.
The SFG and AGN host populations are better isolated from each other in barred galaxies.
Even when leaving the AGN phase and entering the quiescent phase, 
barred galaxies evolve more quickly.

The bar effect in the right panel shows the features more clearly.  
The bar effects are positive overall. 
The $\fagn$ in barred galaxies is higher than that in non-barred galaxies across overall space, 
which is consistent with previous studies \citep{Oh2012, Galloway2015}. 

The most significant bar effect on AGN activity occurs when leaving the SF sequence.
The bars are a great help for AGN triggering in many galaxies with a smaller $\sigma$
(i.e., less massive BHs), accelerating galaxy evolution;
while many of the non-barred counterparts seem to be slowly quenched 
by simple depletion of gas supply through SF
without AGN triggering due to the difficulty in gas transport into
the nuclear region. 
In contrast, in galaxies with a larger $\sigma$, 
the bar effect on AGN triggering is relatively smallest but the $\fagn$ is largest.
Consequently, all AGN occurrences and rapid SF quenching
are closely related to each other, which
requires a massive BH or an efficient gas transport mechanism 
into the nuclear region such as bars.

There is another notable feature that 
AGN hosts are most frequently found 
at an almost constant colour of $u-r\sim2.2$, 
regardless of $\sigma$ and bar presence or absence. 
Among the non-barred galaxies with a small $\sigma$, a small fraction 
experience the AGN phase, but their $\fagn$ peaks at about $u-r = 2.2$, 
as with that of the barred counterparts.
We infer that in our sample galaxies with $\sigma>70\kms$ and $M_{\rm r}<-19.5$,
once an AGN is triggered by any quenching event, the host galaxy quickly evolves into a green valley.

Considering that gas outflows are prevalent among local Type II AGNs
\citep{Woo2016},  
our findings support a negative AGN feedback scenario in which 
outflows from the AGN can push away cold gas, hindering 
SF activity in the host galaxy 
\citep{Silk1998, Dimatteo2005, Hopkins2006, Scannapieco2012}.

We again find the positive bar effect on AGN triggering
in low-$\SFRf$ galaxies having a red colour of $u-r>2.4$.
The quite sudden SF quenching led by bars is seen in
the left panel in Figure~\ref{fig:fig1} above, and the $\fagn$ increase
led by bars is seen in the right panel in Figure~\ref{fig:fig2}.
Even in this case of galaxies lacking cold gas fuel, 
Bars can produce rapid SF quenching and BH feeding. 

\begin{figure*}
    \centering   
   \includegraphics[scale=0.9]{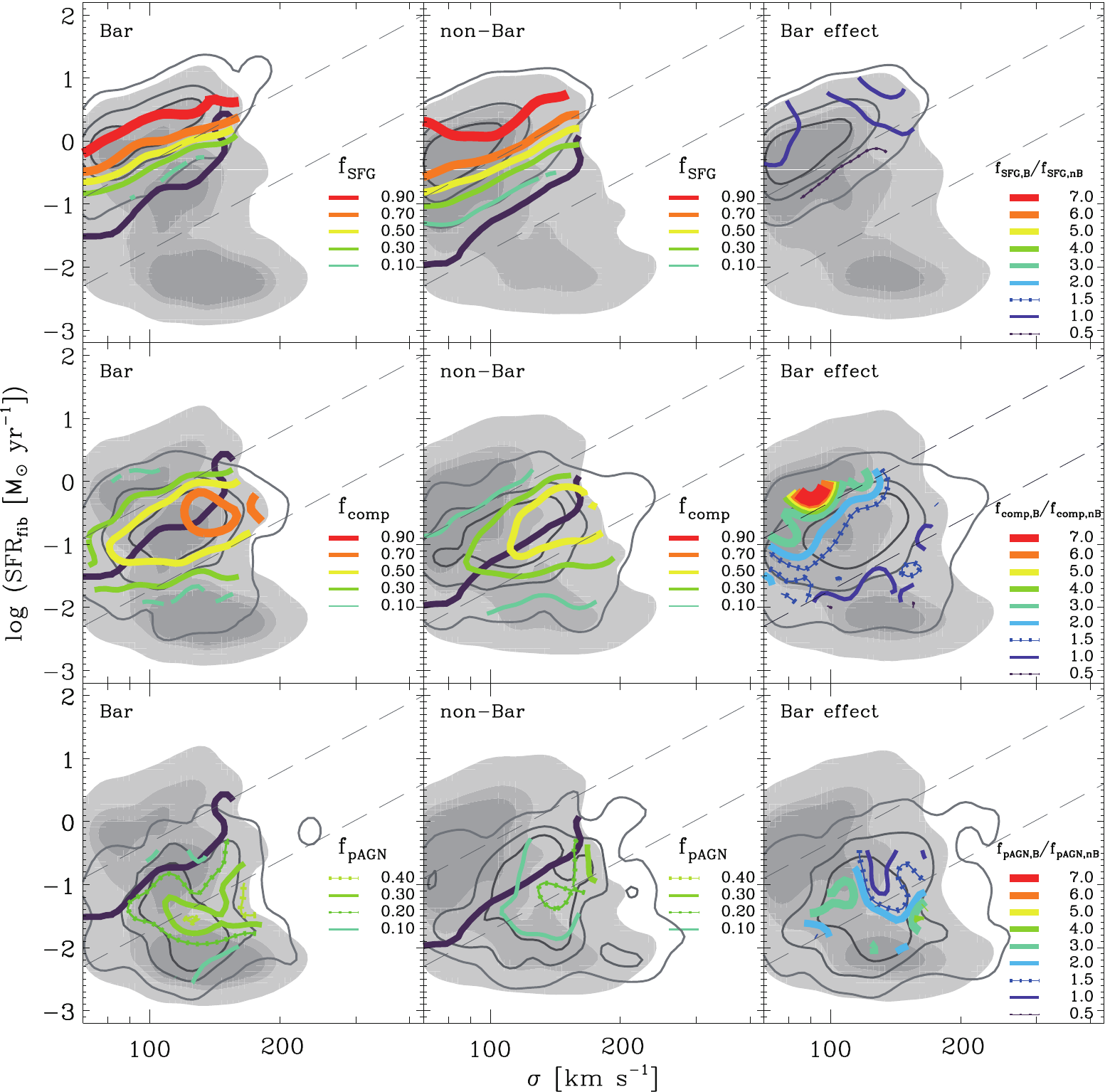}
    \caption{Distributions of bar effect in $\SFRfib$-$\sigma$ space 
for different spectral classes of starburst(top), 
composite (middle), and pure-AGN (bottom). 
}
    \label{fig:fig3}
\end{figure*}

Meanwhile, 
we divided galaxies into the spectral classes of 
SFGs, starburst-AGN composites, and pure AGNs, which are 
expected to be at different evolutionary stages.
We examine how bars affects the evolutionary stages;
the result is shown in Figure~\ref{fig:fig3}.
Spectral classes of SFG, starburst-AGN composite, and pure-AGN are shown 
in the top, middle, and bottom panels, separately. 

The SFG fraction $\fSFG$ has almost the same dependency to that of the composite
fraction $f_{\rm comp}$, and as $\fSFG$ decreases, $f_{\rm comp}$ increases,
suggesting that AGN detection occurs along with central SF quenching. 
The sudden and rapid transitions from SFGs 
to the starburst-AGN composite hosts or to pure-AGNs are produced effectively by bars, 
which is obviously seen at smaller $\sigma$s.
At low $\SFRf$, 
there is a significant positive bar effect on pure-AGN triggering. 

These results are contrary to the result of \citet{Lee2012b} 
that pure-AGN activity is not enhanced 
by strong bars at fixed $u-r$ colour and $\sigma$.
The only difference between this study and Lee et al. is that 
they examined total $u-r$ colour instead of central SFR, which means that
the central SFR is more tightly coupled to AGN activity 
than the global properties of a galaxy, such as total SFR and $u-r$ colour.
The stronger link between AGN triggering and central SF better 
reveals the role of strong bars. 
\citet{Hopkins2012} argued that negative AGN feedback actually acts on
only a very small fraction of remaining gas.
\citet{Karouzos2016} performed a spatially resolved kinematic analysis
of moderate luminous type II AGNs using Gemini Multi-Object Spectrograph IFU data and
claimed that negative feedback is confined to the central kpc region of a galaxy. 

\subsubsection{Bar Effect on AGN Power}
We next ask how the bar affect AGN outflow itself?
We expect that, when gas fuel and BH mass are given,
bars induce efficient gas-inflow, leading to stronger outflows.

To trace the outflow signature of AGN and AGN luminosity,
we adopt $\sigma_{\rm[OIII]}$ and $L_{\rm[OIII]}$, respectively.
\citet{Woo2016} argued that the highly ionized [O III] line is a good tracer of 
AGN-driven gas outflow.
Larger [O III] velocity dispersion, $\sigma_{\rm[OIII]}$,
compared to $\sigma$ indicates that the non-gravitational component in
$\sigma_{\rm[OIII]}$ is larger than the gravitational component,
presumably due to a stronger outflow effect.
The $\sigma$ is used as a proxy for the kinetic component in $\sigma_{\rm[OIII]}$
due to bulge gravitational potential.

\begin{figure}
    \centering   
   \includegraphics[scale=0.5]{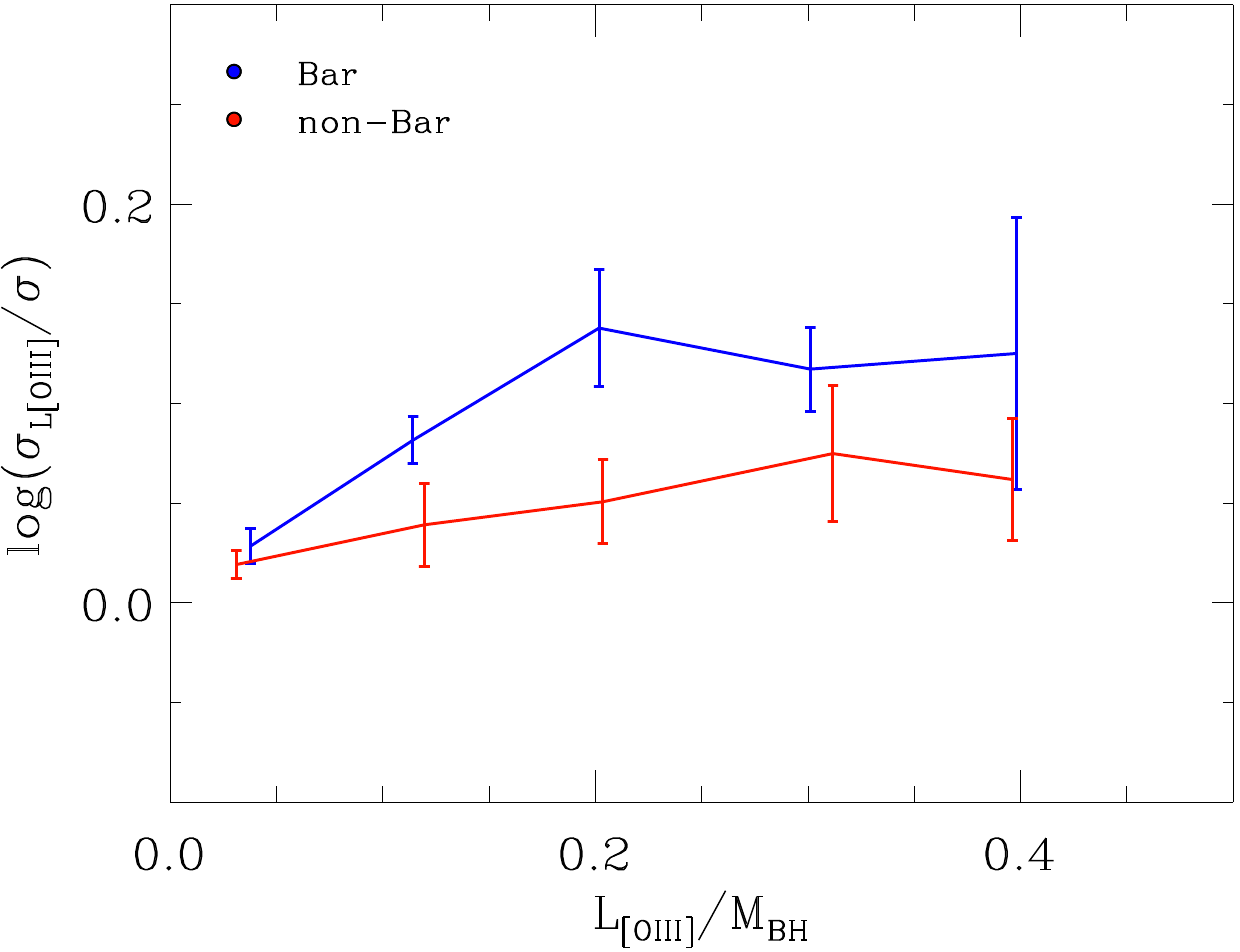}
    \caption{Outflow levels of barred (blue) and non-barred (red) AGN
    hosts as a function of $L_{\rm[O III]}/M_{\rm BH}$. 
The error bars are calculated using the bootstrap method with 1000 runs. 
}
    \label{fig:fig4}
\end{figure}

In Figure~\ref{fig:fig4},
we plot the $\sigma_{\rm[OIII]}/\sigma$ measurements as a function of Eddington ratio 
for barred and non-barred AGN hosts.
The Eddington ratio is defined by $L_{\rm[OIII]}/M_{\rm BH}$.
The $M_{\rm BH}$ is a BH mass 
calculated through the $M_{\mathrm BH} - \sigma$ relation for spiral galaxies, 
as given by \citet{Gultekin2009}.

The average ratio of $\sigma_{\rm[OIII]}/\sigma$ is larger than 1 overall,
indicating that the non-gravitational component in $\sigma_{\rm[OIII]}$ surpasses 
the gravitational component overall.
For a barred case, the effect of the non-gravitational component 
is significantly larger than that for non-barred hosts,
excluding AGNs with lowest $L_{\rm[OIII]}/M_{\rm BH}$.
A positive relation of AGN luminosity and outflow signature is also found.

\begin{figure*}
    \centering   
   \includegraphics[scale=0.9]{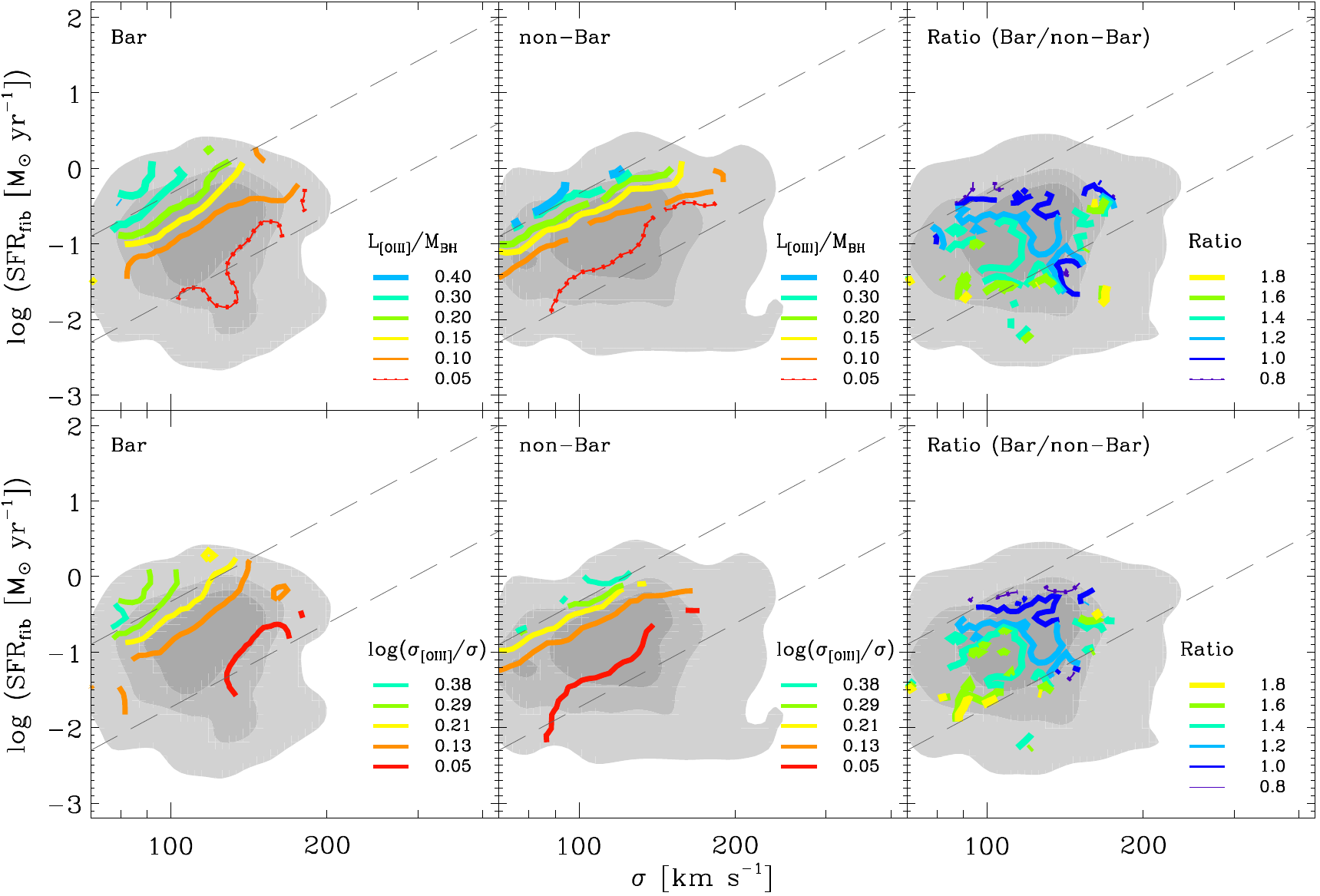}
    \caption{Distributions of $\sigma_{\rm[O III]}/\sigma$ (top) and 
    Eddington ratio $L_\mathrm{[OIII]}/M_\mathrm{BH}$  (lower) of AGN hosts 
    in barred (left) and non-barred (middle) samples in $\SFRfib$-$\sigma$ space. 
    In the left and middle panels, a coloured thick contour denotes a constant Eddington ratio
    or $\sigma_{\rm[O III]}/\sigma$. 
    The coloured contours in each right panel represent the ratio of the measurements 
    for barred and non-barred AGN hosts.
}
    \label{fig:fig5}
\end{figure*}

In Figure~\ref{fig:fig5}, we plot the distributions of $L_{\rm[OIII]}/M_{\rm BH}$ (upper) and 
$\sigma_{\rm[OIII]}/\sigma$ (lower) on the $\SFRf$ and $\sigma$ space.
For comparison, the ratios for the barred and non-barred AGN hosts are plotted in the right panels. 
The uncertainty of each measurement is calculated by 1000 bootstrap resamplings.
The contours are plotted for bins with relative uncertainty less than 1/3.
Each right panel represents the ratio of measurements for barred and non-barred AGN hosts. 

The two distributions are very similar.
At a given $\sigma$, as $\SFRf$ increases, 
both $L_{\rm[OIII]}$ and $\sigma_{\rm[OIII]}$ also increase.
The increase in $\sigma_{\rm[OIII]}$ indicates an increase in the 
effect of its non-gravitational component.

The right panels show that, 
since AGNs are rarely observed at the beginning stage of the evolution from starbursts to AGNs, 
it is difficult to compare the difference in $L_{\rm[O III]}$ (or $\sigma_{\rm[OIII]}$) 
of barred and non-barred AGN hosts 
However, in the intermediate $\SFRfib$ region where most AGNs are observable, 
the Eddington rate and outflow effect of the barred AGN hosts are 
obviously higher than those of non-barred counterparts overall.
Meanwhile, in galaxies with larger $\sigma$ and higher $\SFRf$,
the positive bar effect on the power of AGNs is not noticeable,
which is consistent with the results of Figure~\ref{fig:fig2} of 
the smallest positive bar effect on AGN triggering in these galaxies.

On the other hand, when most AGN hosts are observed at intermediate $\SFRf$s,
both their $L_{\rm[O III]}$ and $\sigma_{\rm[O III]}$ 
have been significantly reduced at a given $\sigma$, 
which is consistent with the result of \citep{Woo2017}. 
This suggests that
the negative feedback from observable AGNs today
has already declined significantly. 

\begin{figure*}
    \centering   
   \includegraphics[scale=0.9]{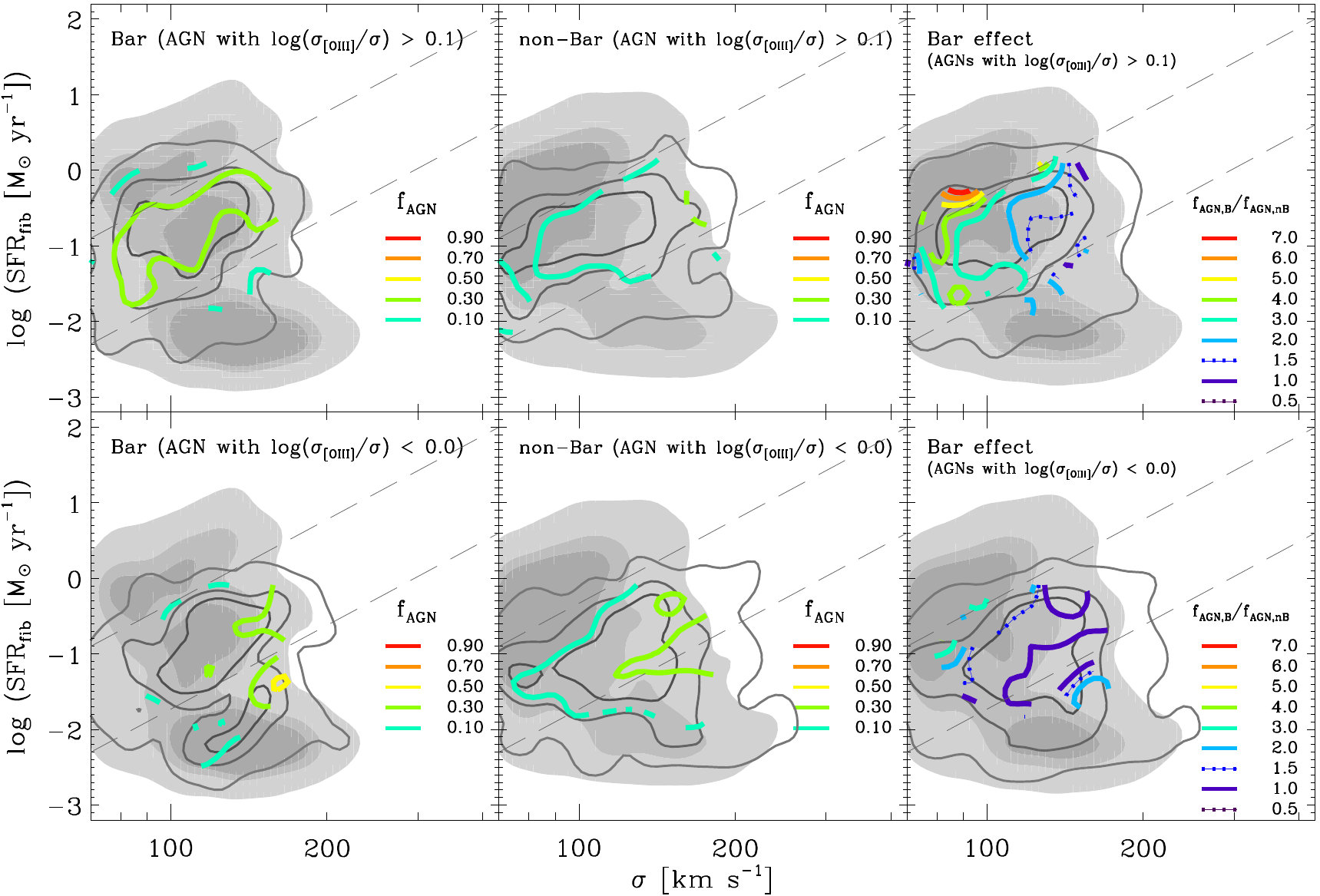}
    \caption{Difference in the bar effect on AGN activity with two strengths of outflows 
    for $\sigma_{\rm[O III]}\ge 1.25 \sigma$ (top) and $\sigma_{\rm[O III]}<\sigma$ (bottom). 
    Distributions of AGN fraction of barred (left) and non-bared (middle) galaxies, 
    and the bar effect on AGN activity (right) on the space are plotted for the two cases.}
    \label{fig:fig6}
\end{figure*}

To verify the positive role of bars, we examine the bar effect on the strength of AGN outflow.
We divide the AGNs into two cases with and without strong outflow signature
and compare the bar effects on AGN power, as in Figure~\ref{fig:fig6}. 
The results clearly demonstrate that when $\SFRf$ and $\sigma$ are fixed, 
bars in small-$\sigma$ galaxies play a critical role in triggering powerful AGNs with 
a strong outflow with $\sigma_{\rm[O III]}\ge 1.25\sigma$;
in triggering weak AGNs with $\sigma_{\rm[O III]}<\sigma$ in the galaxies,
the role of bars is significantly reduced.

\subsection{Bar Effect and Bulge Type}\label{sec:CMC}
The bar-driven gas supply contributes to some fraction of CMC,
but the CMC growth also rather weakens or dissolves the bars
\citep{Hasan1990, Pfenniger1990, Friedli1993}. 
Several studies argued that 
the gravitational forcing of massive bulge slows bar formation 
\citep{Athanassoula2013, Cheung2013}.

Motivated by these results, we examine 
how build-up of bulge and bar structure are related and 
how a well-developed bulge affects galaxy evolution.

First, we compare how bar fraction depends on bulge type.
For convenience, to distinguish classical bulge from pseudo-bulge in our sample,
we adopt 
the Petrosian concentration index with a cut of $C=2.6$ that was used in \citet{Strateva2001} 
to separate ellipticals and spiral galaxies.
The $C$, is defined as the ratio between central radii of 
a galaxy containing 50\% and 90\% of the $i$-band Petrosian flux, $R_{90}/R_{50}$.
\citet{Gadotti2009} argued that the light concentration index is a better proxy for the bulge-to-total ratio 
than the global Sersic index, and that galaxies with $C\ge2.6$ include only few pseudo-bulges.

\begin{figure}
    \centering   
   \includegraphics[scale=0.6]{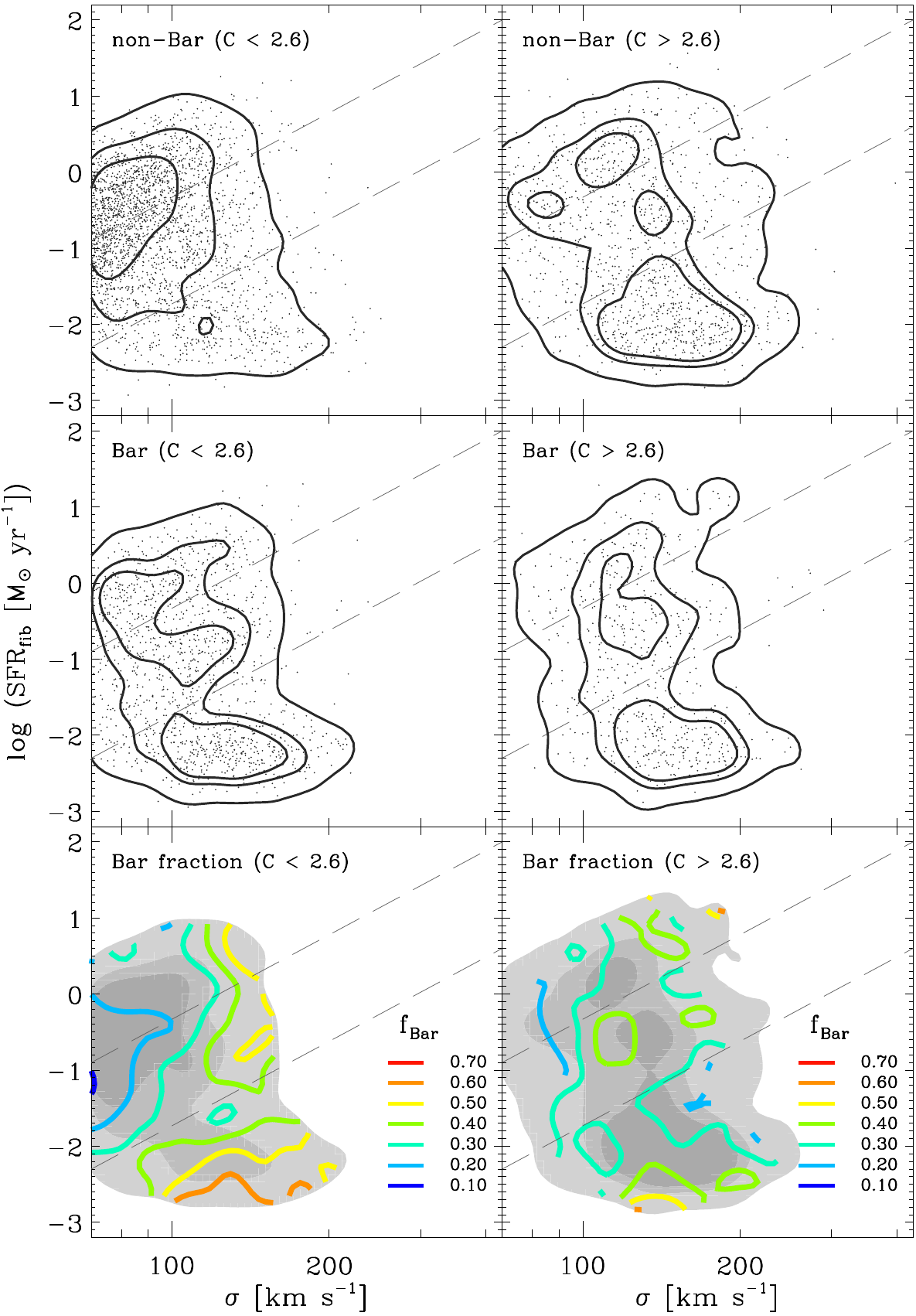}
    \caption{Distributions of the bar fraction for the pseudo- (bottom left)
and classical bulge sample (bottom right).
The coloured thick solid line represent constant fractions and 
grey thin contours or grey filled line contours represent 
constant number densities of each sample, respectively.
}
    \label{fig:fig7}
\end{figure}

The results are shown in Figure~\ref{fig:fig7}.
The bottom panels show the distributions of $\fbar$ for pseudo- and classical bulge samples.
Interestingly, the overall bar fractions for both bulge types are 0.3.
Majority of the pseudo-bulge galaxies remain in the main sequence of SFGs and
rarely have bars (see the bottom left panel). 
They also have a strong dependency of $\fbar$ on $\sigma$ and $\SFRfib$.
More bars found among quenched galaxies indicate that
bars contribute to pseudo-bulge galaxy evolution.

In contrast,
in classical bulges, the dependency of $\fbar$ on the $\sigma$ and $\SFRfib$ is weak
and many of the non-barred cases, as well as the barred cases have already been quenched
(see right panels). 
Some of non-barred cases have larger $\sigma$ than barred case before evolving to the quiescent stage, 
indicating that they assembled the concentrated bulge component early
in evolution, without the help of bars.
The result is not surprising 
because classical bulges are thought to be formed by major galaxy mergers \citep{Kormendy2016}.
Consequently, 
these results clearly show 
that classical bulge galaxies require alternative quenching pathways such as major mergers
\citep[e.g.,][]{Toomre1972, Heckman1986, Hopkins2006}
and galaxy-galaxy interactions \citep{Goulding2018}.

To support this conclusion, 
we examine the difference of bar effect on AGN between the two bulge-type cases.
One can expect that the bar effect in classical bulges is not as much as in pseudo-bulges.  
\begin{figure*}
    \centering   
   \includegraphics[scale=0.9]{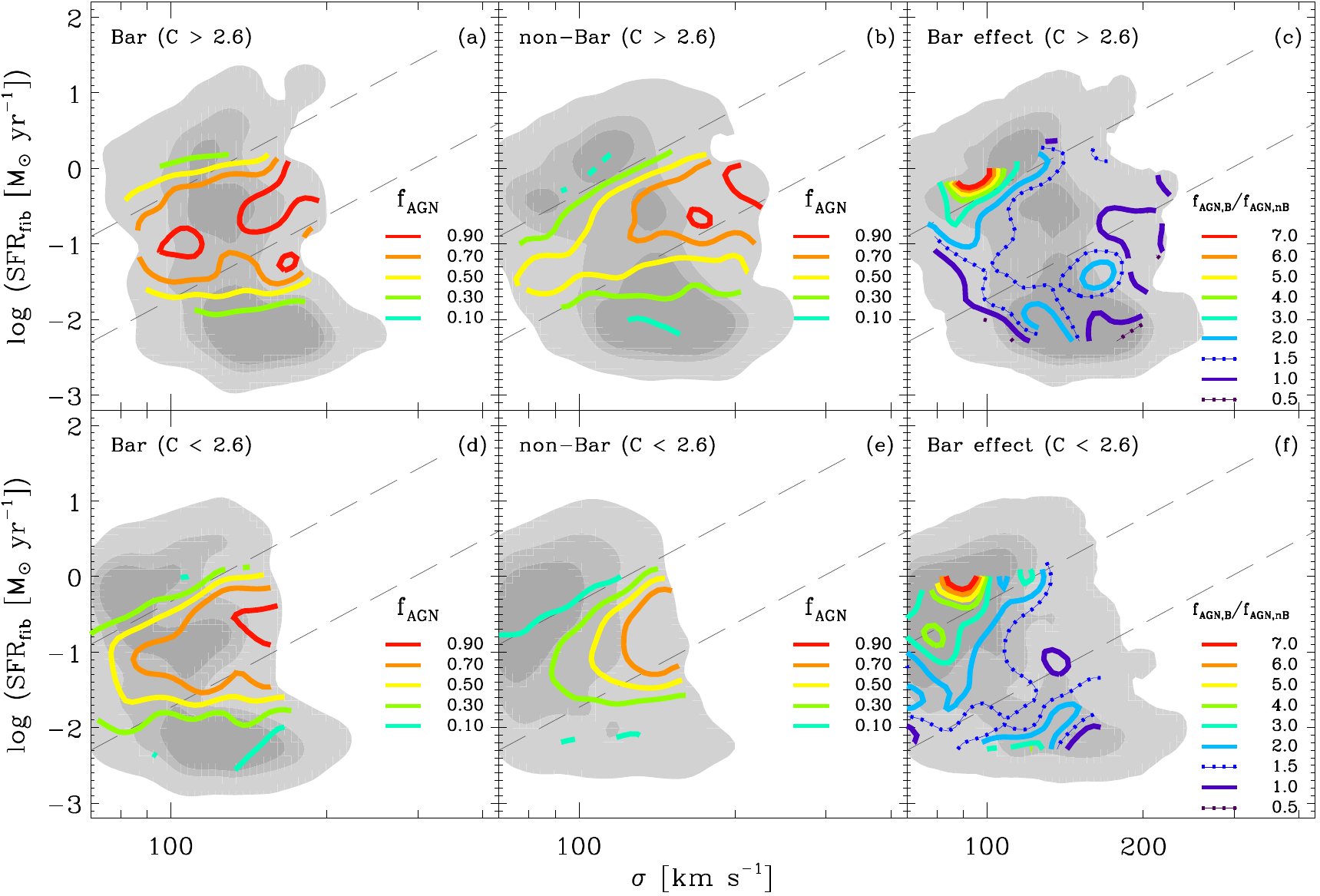}
    \caption{Distributions of AGN fraction of barred (left) and 
non-bared (middle) galaxies on the $\SFRfib$-$\sigma$ space. 
The distributions are plotted separately for galaxies with classical-bulge
of $C\ge2.6$ (upper) and pseudo-bulge of $C<2.6$ (lower).
The bar effect on AGN activity is plotted at the right panels.
}
    \label{fig:fig8}
\end{figure*}
The result is shown in Figure~\ref{fig:fig8}. 
At a given bulge-type, the bar effect on $\fagn$ is shown in the right panels.
Independent of bulge-type, a certain positive bar effect on AGN triggering
is found over the entire space.

The most significant difference between the bulge-types 
is found in intermediate-$\SFRfib$ galaxies.
The results from panels (c) and (f) in Figure~\ref{fig:fig8}
demonstrate that bar-driven evolution from starburst to AGN
is more effective in pseudo-bulge cases, particularly ones with a lower $\sigma$.

That can be explained by 
the classical-bulge tendency to interrupt the bar-driven gas inflow into BH
\citep[see,][]{Norman1996}. 
However, the explanation is ruled out because, in barred galaxies,
the classical bulge case has higher $\fagn$ values than the pseudo-bulge case
(see the panels (a) and (d)),
which is also found in non-barred galaxies (see panels (b) and (e)). 
The result clearly shows that 
well-developed bulges are helpful for AGN triggering 
in galaxies with a lower $\sigma$.
 
The lower bar effect found in classic bulge galaxies 
is because AGN triggering in the non-barred case is 
relatively easy compared to that in the pseudo-bulge counterparts
(see panels (b) and (e)).

\subsection{Scaled Bar Lengths}\label{sec:length}
Many previous studies have showed that bar length varies with galaxy properties; 
for example, bars in early-type galaxies are clearly longer than bars in late-type galaxies
\citep[e.g.,][]{Elmegreen1996, Erwin2005, Laurikainen2017}. 
The relative bar length of star-forming galaxies increases with increasing $\sigma$
and is longer at higher $\sigma$ than that of AGN hosts \citep{Oh2012}. 
Galaxies with classical-bulges tend to have longer bars than ones with pseudo-bulges, 
and relative bar lengths of galaxies with star-forming classical bulges are largest
\citep{Cheung2013}. 

\begin{figure*}
    \centering   
   \includegraphics[scale=0.4]{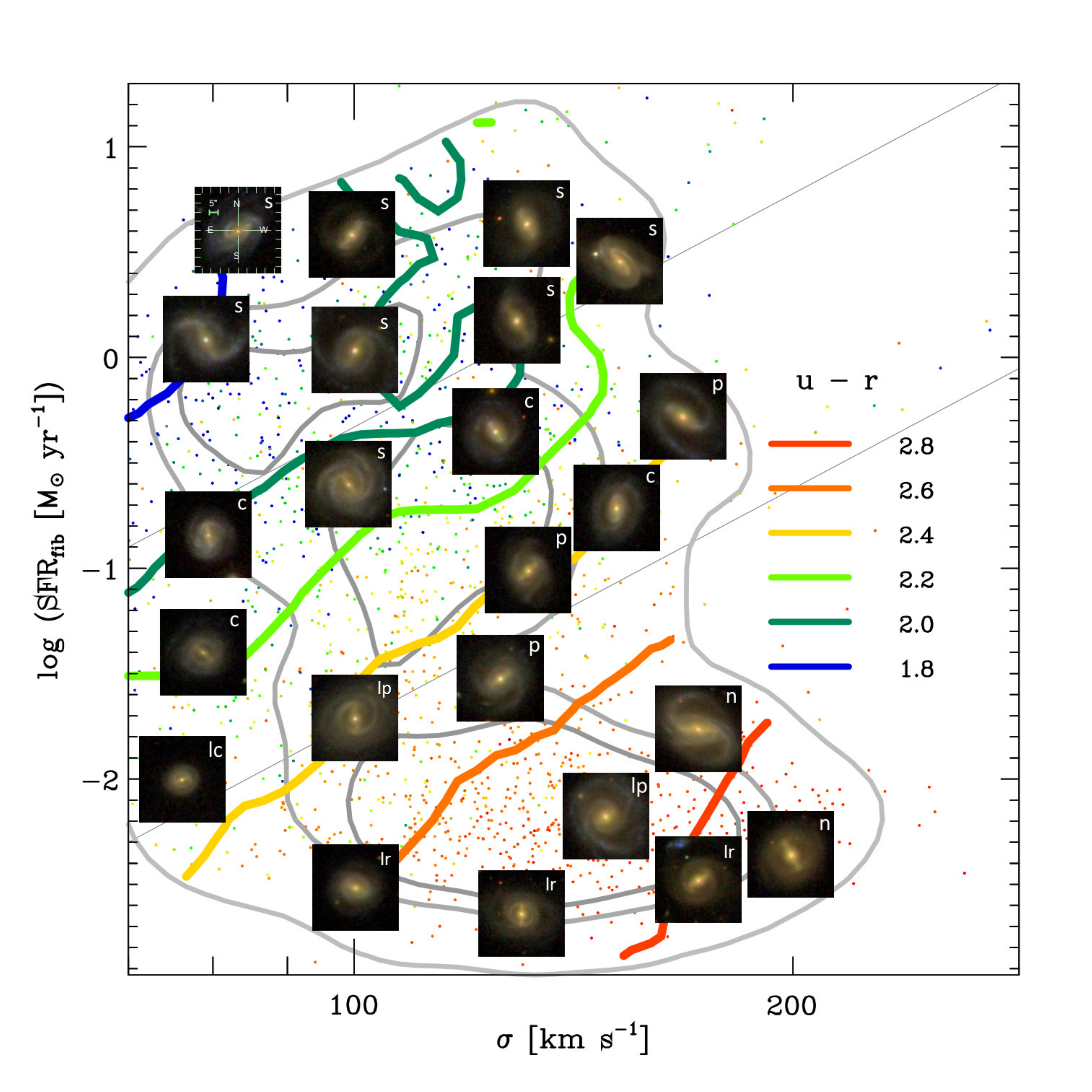}
    \caption{Sample SDSS colour-composite images for some of our barred 
galaxies plotted on the same parameter space.
Each image covers 48$\arcsec \times 48\arcsec$.
The abbreviations on the image represent the spectral type of galaxy:
s, c, p, lp, lr, and n consistent with SFG, composite galaxy,  
pure AGN, low-$S/N$ pure AGN, low-$S/N$ retired galaxy, and non-AGN, respectively. 
}
    \label{fig:fig9}
\end{figure*}

In this section, we explore how bar length is related to evolutionary sequence.
First, we list SDSS colour-composite images of 
some sample galaxies with bar in the range of $0.03<z<0.04$ in Figure~\ref{fig:fig9}.
The SDSS images show how much  
the area corresponding the finite size of the optical fibre ($3''$)
covers the inner region of our sample galaxies.
Before looking at the results, it should be kept in mind that 
the galaxies in our sample
have relatively large $\sigma$ values of $70\kms$ or more, 
so many disk-dominated and faint spiral galaxies possibly are excluded,
and only strong bars are included in our bar selection.

As in Figure~\ref{fig:fig9}, 
it is difficult to distinguish the relationship between bar length and 
morphological characteristics of its host galaxy 
by visual inspection.
Nevertheless, we find a notable feature 
that, at a given $\sigma$, bars in galaxies with higher $\SFRfib$ are longer.

Therefore, we adopt the bar length measurements provided by a galaxy zoo project
\citep{Hoyle2011}. 
Out of 1,893 barred galaxies in our sample, 1,102 are matched
with the galaxy zoo sample.
Most of the missing galaxies are those with 
apparent $r$-band magnitudes greater than 17.0.
Bar length is normalized to two times the $r-$band Petrosian radius 90 of the galaxy.
In Table~\ref{tab:tab3}, we list the mean values for 
the relative bar length for six subsamples with different $\SFRfib$ and $\sigma$.
We also list the values when the morphological type of galaxy bulge is fixed.

Barred galaxies with large $\sigma$s tend to have longer bars overall.
When a bulge-type is fixed, the $\sigma$ dependency of the scaled bar length mostly disappears,
indicating that bar length is closely related to concentration.
It is clear that concentrated bulge galaxies have longer bars than less-concentrated bulge ones.
At a given $\sigma$, bar length decreases as $\SFRfib$ decreases,
which is consistent with the results of previous observational studies
\citep{Martinet1997, Cheung2013}.
\citet{Oh2012} gave a similar result that, at a given large $\sigma$, 
the bars in galaxies with high central SFR are longer.
It is noteworthy that the low-$\SFRfib$ galaxies tend to have much shorter bars than other galaxies.
Our observational findings demonstrate that a bar not only accelerates the evolution of host
galaxy, but a bar itself also evolves along with the galaxy evolution.

\section{Summary and Conclusion}\label{sec:discuss}
We explored the significance of bars in AGN activities and star formation quenching
in galaxies using a volume-limited face-on late-type galaxy sample with $M_r<-19.5$, $\sigma>70\kms$,
and $0.02<z<0.054$ selected from SDSS DR7.
Out of 6,195 galaxies, 1,659 Type II AGNs are conservatively identified, 
and 1,893 strong-barred galaxies are classified as barred galaxies
using the bar classification visually inspected by \citet{Lee2012b}.
Our barred galaxy sample is biased towards early-type 
morphology because of the lower cut of $\sigma$.
The redshift range was determined so that the spectroscopic fibre aperture could enclose
the bulge at a galactic center. 

We found a distinct trimodal distribution of galaxies in $\SFRfib$ and $\sigma$ space
suggesting that SF quenching 
occurs through two short-lived transition processes, which is particularly seen in barred galaxies. 

The AGN fraction distribution on the same space shows that 
the AGN fraction increases with decreasing SFG fraction and 
peaks at the intermediate $\SFRfib$ region, suggesting that
SF quenching and AGN occurrence are accompanied.
Outcomes from barred galaxies support this scenario, 
showing more sudden quenching and higher AGN fraction. 
This also indicates that bars are very capable of transporting 
cold gas fuel into the nuclear region.

Here, when most AGNs are observed behind the SF sequence, 
their Eddington ratio and outflow strength have already been reduced.
In other words, 
the two phenomena of central SF and AGN activity take place 
together without a large delay. 
 In this context, we postulate that a low AGN fraction at high $\SFRf$ is 
due to the difficulty in observing the triggered AGNs in this optical survey
rather than the difficulty in inducing BH accretion itself.
The dense dusty gases surrounding 
the nuclear region obscure the AGNs 
\citep[see section 3.3 of ][for a review]{Hickox2018}.
\citet{Chen2015} found that the obscured quasar fraction 
among far-IR quasars increases significantly with far-IR luminosity increasing.

Our findings allow us to infer that powerful gas outflows driven by AGNs lead to 
rapid galaxy quenching.
The gas outflows, which are prevalent among local Type II AGNs \citep{Woo2016},
push out the gas and dust enshrouding a BH and
negate future gas fuel supply to the BH, resulting in SF quenching and AGN unveiling. 

When most AGNs are detected, 
the role of bars as an inducer of AGN activity is markedly large when
galaxies have smaller $\sigma$s, i.e., less massive BHs.
The positive role is also prominent even in galaxies lacking cold gas fuel,
even in galaxies with massive BHs.

Galaxies without a bar, particularly those having a less massive BH, 
seem to be slowly quenched by simple depletion of gas supply through SF,
resulting in relatively smooth transition during the evolutionary sequence.
A smaller fraction of galaxies experience the AGN phase which is weak.

In conclusion, the abundant gas fuel and massive BH are paramount to galaxy evolution.
If these two sufficient conditions are satisfied, galaxies evolve quickly
without great help from bars.
It is obvious that bar instability accelerates the entire process of galaxy evolution
by facilitating gas consumption through SF and gas inflow into BH,
which, however, is greatly helpful in galaxies where either of the two conditions is not sufficient.

In addition to bar instability, 
external processes such as continuous interactions, minor merger \citep[e.g.,][]{Sancisi2008}, tidal interaction, and so on 
are responsible for another effective gas transport mechanism. 
This provides a significant source of fuel for gas accretion to maintain long-lived bars or reform bars
\citep{Bournaud2002}. 
The bar reformation can be also triggered by tidal interaction \citep{Gerin1990, Berentzen2004} 
or a strong tidal field in a cluster \citep{Byrd1990}.
In a future paper, 
we will assess the importance of the external process 
on galaxy evolution in addition to bar instability. 

\section*{Acknowledgements}

We acknowledge support from the National Research Foundation (NRF) of 
Korea to the Center for Galaxy Evolution Research (No. 2017R1A5A1070354).  
The work by S.S.K. was supported by the NRF grant 
funded by the Ministry of Science and ICT of Korea (NRF-2014R1A2A1A11052367). 
This work was also supported by the BK21 plus program through the NRF 
funded by the Ministry of Education of Korea. 

Funding for the SDSS and SDSS-II has been provided by the Alfred P. Sloan Foundation, the Participating Institutions, the National Science Foundation, the U.S. Department of Energy, the National Aeronautics and Space Administration, the Japanese Monbukagakusho, the Max Planck Society, and the Higher Education Funding Council for England. The SDSS Web site is http://www.sdss.org/.
The SDSS is managed by the Astrophysical Research Consortium for the Participating Institutions. The Participating Institutions are the American Museum of Natural History, Astrophysical Institute Potsdam, University of Basel, Cambridge University, Case Western Reserve University, University of Chicago, Drexel University, Fermilab, the Institute for Advanced Study, the Japan Participation Group, Johns Hopkins University, the Joint Institute for Nuclear Astrophysics, the Kavli Institute for Particle Astrophysics and Cosmology, the Korean Scientist Group, the Chinese Academy of Sciences (LAMOST), Los Alamos National Laboratory, the Max-Planck-Institute for Astronomy (MPIA), the Max-Planck-Institute for Astrophysics (MPA), New Mexico State University, Ohio State University, University of Pittsburgh, University of Portsmouth, Princeton University, the United States Naval Observatory, and the University of Washington.






\begin{table*}
   \centering
   \caption{Spectral Type Classification of Sample based on Different S/N cut of Emission lines}
   \label{tab:tab1}
   \begin{tabular}{lrr} 
      \hline
      \hline      
       Number (Fraction) & ${\rm S/N} \geq 6$ & ${\rm S/N} \geq 3$ \\
      \hline
         All                  &  6195  (1.00)  &  6195 (1.00) \\
         SFG                  &  1943  (0.31)  &  2507 (0.40) \\
         Total AGN            &  1653  (0.27)  &  2107 (0.34) \\
         ~~~~~~~~~~Composite  &  1190  (0.19)  &  1584 (0.26) \\
         ~~~~~~~~~~Pure AGN   &   463  (0.07)  &   523 (0.08) \\
         Retired galaxy$^{a}$ &   261  (0.04)  &   685 (0.11) \\
      \hline
      \multicolumn{3}{l}{Note: $^{a}$ Retired galaxies are powered by the hot low-mass evolved stars rather than by low luminosity AGNs.}\\
   \end{tabular}
\end{table*}

\begin{table*}
   \centering
   \caption{Sample Statistics for Each Subsample}
   \label{tab:tab2}
   \begin{tabular}{lccc} 
      \hline
      \hline      
       Classification (Number) & All (6195) & Bar (1893) & non-Bar(3754) \\
      \hline
      \hline
       High Central SF Sample \\     
      \hline
         All        &  100\%  (1952)  &  100\% (473)  &  100\% (1330) \\
         SFG        &   83\%  (1612)  &   82\% (388)  &   83\% (1108) \\
         Composite  &    5\%  ( 102)  &   10\% ( 47)  &    4\% (  46) \\
         Pure AGN   &    1\%  (  19)  &    1\% (  5)  &    1\% (  14) \\
      \hline
       Intermediate Central SF Sample \\
      \hline
         All        &  100\%  (2283)  &  100\% (662)  &  100\% (1428) \\
         SFG        &   14\%  ( 328)  &    7\% ( 45)  &   17\% ( 247) \\
         Composite  &   39\%  ( 896)  &   55\% (362)  &   32\% ( 464) \\
         Pure AGN   &   10\%  ( 233)  &   17\% (112)  &    7\% (  99) \\
      \hline
       Low Central SF Sample \\
      \hline
         All        &  100\%  (1960)  &  100\% (758)  &  100\% (996) \\
         SFG        &    0\%  (   3)  &    0\% (  0)  &    0\% (  3) \\
         Composite  &   10\%  ( 192)  &    8\% ( 57)  &   11\% (110) \\
         Pure AGN   &   11\%  ( 211)  &   13\% ( 97)  &   10\% ( 96) \\
      \hline
      \multicolumn{3}{l}{Note: The three subsamples are divided by two diagonal lines plotted in Fig.~\ref{fig:fig1}.}\\
   \end{tabular}
\end{table*}

\begin{table*}
   \centering
   \caption{Mean Values of Relative Bar Length for Each Subsample}
   \label{tab:tab3}
   \begin{tabular}{rrr} 
      \hline
      \hline      
       Subsample (Number) & $70 < \sigma < 130 \kms$ & $\sigma > 130 \kms$ \\
      \hline
      \hline
       All Galaxies \\     
      \hline
         High $\SFRfib$ (243)      &  $0.51\pm{0.01}$ (205) & $0.54\pm{0.02}$  (38)   \\
		 Intermediate $\SFRfib$ (338) &  $0.47\pm{0.01}$ (256) & $0.50\pm{0.01}$  (82)   \\ 
         Low $\SFRfib$ (521)          &  $0.40\pm{0.01}$ (225) & $0.43\pm{0.01}$ (296)   \\
      \hline
       Galaxies with $C\ge2.6$ \\
      \hline
		 High $\SFRfib$ (71)         &  $0.57\pm{0.02}$  (53) & $0.60\pm{0.03}$  (18)   \\
		 Intermediate $\SFRfib$ (95) &  $0.53\pm{0.02}$  (56) & $0.56\pm{0.02}$  (39)   \\ 
		 Low $\SFRfib$ (159)         &  $0.43\pm{0.02}$  (35) & $0.46\pm{0.01}$ (124)   \\
      \hline
       Galaxies with $C<2.6$ \\
      \hline
		 High $\SFRfib$ (172)         &  $0.50\pm{0.01}$ (152) & $0.50\pm{0.03}$  (20)   \\
		 Intermediate $\SFRfib$ (243) &  $0.46\pm{0.01}$ (200) & $0.48\pm{0.02}$  (43)   \\ 
		 Low $\SFRfib$ (362)          &  $0.40\pm{0.01}$ (190) & $0.41\pm{0.01}$ (172)   \\
      \hline
      \multicolumn{3}{l}{Note: The bar length is normalized to two times the $r$-band Petrosian radius 90.
      A standard error of the mean is also listed.}\\
   \end{tabular}
\end{table*}



\bsp	
\label{lastpage}

\end{document}